\documentclass[twocolumn,amsmath,amssymb,aps]{revtex4}



\usepackage{graphicx}
\usepackage{dcolumn}

\newcommand\mean[1]{\ensuremath{\langle#1\rangle}}

\begin{document}
\title{Information diffusion epidemics in social networks}

\author{Jos\'e Luis Iribarren}
%\email[]{iribarren@es.ibm.com}
\affiliation{IBM Corporation, ibm.com e-Relationship Marketing
Europe, E-28002, Madrid, Spain}

\author{Esteban Moro}
%\email[]{}
\affiliation{Grupo Interdisciplinar de Sistemas Complejos (GISC)
and Departamento de Matem\'aticas, Universidad Carlos III de
Madrid, E-28911, Legan\'es (Madrid), Spain}

\date{\today}

\maketitle

{\bf {\small Abstract:} The dynamics of information dissemination in
social networks is of paramount importance in processes such as
rumors or fads propagation \cite{yamir}, spread of product
innovations \cite{valente} or "word-of-mouth" communications
\cite{womma,buzzbuzz}. Due to the difficulty in tracking a specific
information when it is transmitted by people, most understanding of
information spreading in social networks comes from models
\cite{golden} or indirect measurements \cite{motion}. Here we
present an integrated experimental and theoretical framework to
understand and quantitatively predict how and when information
spreads over social networks. Using data collected in Viral
Marketing campaigns \cite{jurvetson} that reached over 31,000
individuals in eleven European markets, we show the large degree of
variability of the participants' actions, despite them being
confronted with the common task of receiving and forwarding the same
piece of information. Specifically we observe large heterogeneity in
both the number of recommendations made by individuals and of the
time they take to transmit the information. Both have a profound
effect on information diffusion: Firstly, most of the transmission
takes place due to super-spreading events which would be considered
extraordinary in population-average models. Secondly, due to the
different way individuals schedule information transmission
\cite{barabasinature,telephone,blogs} we observe a slowing down of
the spreading of information in social networks that happens in
logarithmic time. Quantitative description of the experiments is
possible through an stochastic branching process \cite{branching}
which corroborates the importance of heterogeneity. The fact that
both the intensity and frequency of human responses show also large
degrees of heterogeneity in many other activities
\cite{pitkow,tipping,sexual} suggests that our findings are
pertinent to many other human driven diffusion processes like
rumors, fads, innovations or news which has important consequences
for organizations management, communications, marketing or
electronic social communities.}

\medskip

Each day, millions of conversations, e-mails, SMS, blog comments,
instant messages or web pages containing various types of
information are exchanged between people. Humans behave in a viral
fashion, having a natural inclination to share the information so as
to gain reputation, trustworthiness or money. This ``word-of-mouth''
(WOM) dissemination of information through social networks is of
paramount importance in our every day life. For example, WOM is
known to influence purchasing decisions to the extent that 2/3 of
the economy of the United States is driven by WOM recommendations
\cite{buzzbuzz}. But also WOM is important to understand
communication inside organizations, opinion formation in societies
or rumor spreading. Despite its importance,  detailed empirical data
about how humans disseminate information are scarce or indirect
\cite{golden,kleinberg}. Most understanding comes from implementing
models and ideas borrowed from epidemiology on empirical or
synthetic social networks \cite{yamir,motion}. However, unlike virus
spreading, information diffusion depends on the voluntary nature of
humans, has a perceived transmission cost and is only passed by its
host to individuals who may be interested on it
\cite{huberman,flow}. Here we present a large scale experiment
designed to measure and understand the influence of human behavior
on the diffusion of information.

\medskip

\begin{figure}
\centering
\includegraphics[width=9cm,clip=]{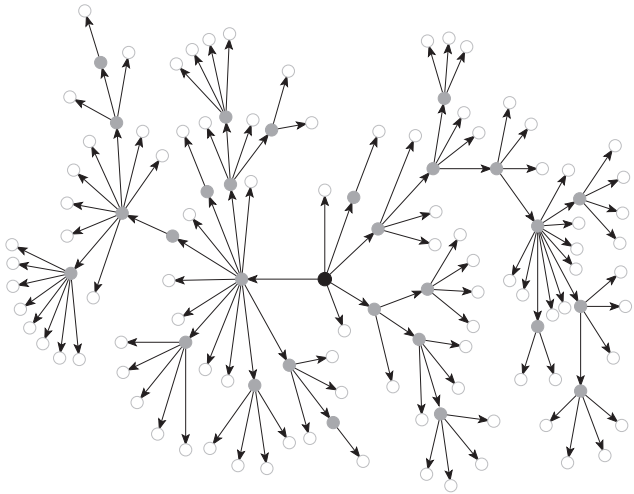}
\caption{ The viral network detected in the
campaigns consists of a large number of disconnected clusters as
this one found in Spain. It has 122 nodes and its diameter (longest
undirected path) is 13. The structure starts out of a seed
participant in the center (black) and grows through secondary viral
propagation of viral nodes (gray) until it reaches this large size.
The probability of finding a similar occurrence in homogeneous
random network models (see Figure \ref{fig2}) is negligible.}
\label{fig1a}
\end{figure}

We analyzed a series of controlled viral marketing \cite{jurvetson}
campaigns in which subscribers to an on-line newsletter were offered
incentives for promoting new subscriptions among friends and
colleagues. This offering was virally spread through recommendation
e-mails sent by participants. This ``recommend-a-friend'' mechanism
was fully conducted electronically and thus could be monitored at
every step. Spurred by exogenous online advertising, a total of
7,153 individuals started recommendation cascades subsequently
fueled through viral propagation carried out by 2,112
\emph{secondary spreaders}. This resulted in another 21,918
individuals touched by the message which they did not pass along
further. All in all, 31,183 individuals were ``infected'' by the
viral message. Of those, 9,265 were spreaders. Thus, 77\% of the
participants were reached by the endogenous WOM viral mechanism. We
call \emph{seed nodes} the individuals spontaneously initiating
recommendation cascades and \emph{viral nodes} the individuals who
pass e-mail invitations along after having received them from other
participants. The topology of the resulting viral recommendations
graph (designated as the Viral Network) is a directed network formed
by 7,188 isolated components, or viral cascades, where nodes
representing participants are connected by arcs representing
recommendation e-mails (see Fig. \ref{fig1a}).

\begin{figure}
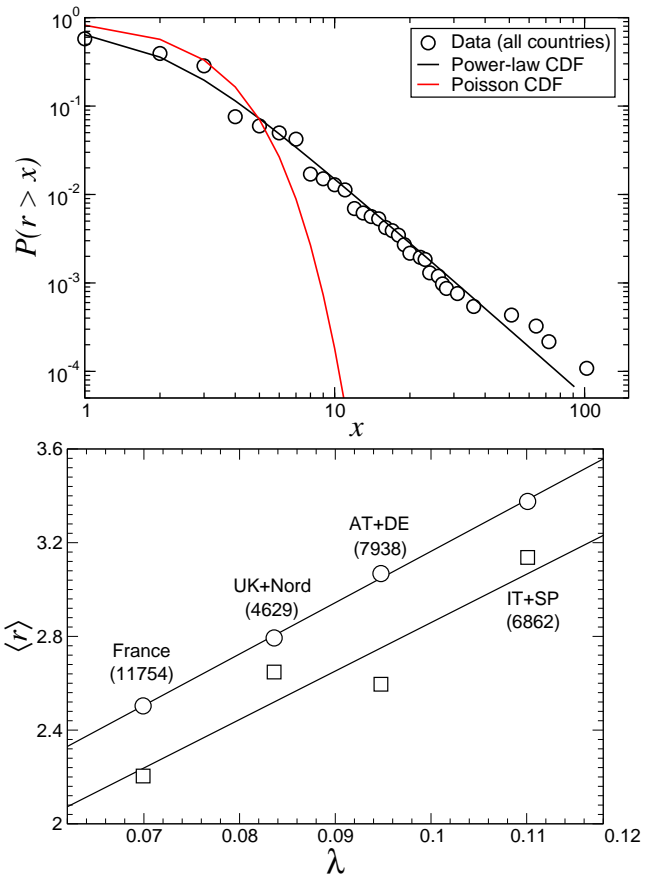

\centering
\includegraphics[width=8.2cm,clip=]{kout.eps}
\includegraphics[width=8.5cm,clip=]{koutlambda.eps}
\caption{Upper panel: Fanout
cumulative probability distribution function for viral campaigns in
all countries (circles). Solid lines show maximum likelihood fits
for power-law $P(r_{v}>x) = H/(\beta+x^\alpha)$ (black circles) with $H$ a normalization constant, and $\beta = 60.07$ and $\alpha = 3.50$ and Poisson probability distribution
functions with mean $\overline{r}_v$ (see appendix \ref{appendixa}). Lower panel: Fanout
Coefficient for viral (circles) and seed (squares) participants as a
function of the Viral Transmissibility $\lambda$ for different
groups of countries. For a given campaign, both parameters are
linearly dependent as $\overline{r}_v=a_v\lambda+b_v$ because the
participants viral decisions stem from evaluating the same utility
function. For the campaigns analyzed the linear fit results in
$a_v=21.9$ and $b_v=0.971$. Variation between countries is due to a
different acceptance of the offering by customers in those markets.}
\label{fig1b}
\end{figure}

\begin{table}
\begin{tabular}{cccccccc}
\hline
Group & Nodes & Cascades & $\overline{r}_{s}$ &
$\overline{r}_{v}$ & $\lambda$ & $\overline{s}$ & $\overline{s}*$
\\ \hline
ALL & 31,183 & 7,188 & \ 2.51 & \ 2.96 &\ 0.088 &\ 4.39 &\ 4.34 \\
SP+IT & 6,862 & 1,162 &\ 3.14 &\ 3.38 &\ 0.11&\ 5.99 &\ 5.91 \\
France & 11,754 & 3,244 &\ 2.20&\ 2.50 &\ 0.070 &\ 3.67 &\ 3.62 \\
AT+DE & 7,938 & 1,743 &\ 2.55 &\ 3.07 &\ 0.095 &\ 4.59 &\ 4.55 \\
UK+Nordic & 4,629 & 1,039 &\ 2.69 &\ 2.79 &\ 0.084 &\ 4.51 &\
4.45\\\hline
\end{tabular}

\caption{The eleven participating countries have been distributed
in four culturally homogeneous groups for statistical relevance.
Network parameters of their corresponding viral network, shown
above, include the theoretical average cascade size $\overline{s}$
predicted by the model through equation ($1$), and the real value
$\overline{s}*$ measured in the campaigns.\label{table1}}
\end{table}

\medskip

The spreading of information or diseases in a population is often
described by average quantities \cite{andersonmay}. Although
infection and propagation can be quite involving processes,
population-level analysis describe viral propagation as a function
of the probability of a virally informed person to become a {\em
secondary spreader} ($\lambda$), and of the average number of people
contacted by {\em secondary spreaders} ($\overline{r}$). Thus, in
this simple approach, two parameters fully characterize the
mean-field description of information diffusion: Viral
Transmissibility ($\lambda$) and Fanout coefficient
($\overline{r}$). In the viral campaigns we found that only $8.79\%$
of the participants receiving a recommendation e-mail engaged in
spreading, and thus $\lambda = 0.0879$. The Fanout coefficient
$\overline{r}$, is the average number of recommendation e-mails sent
by spreading nodes. Its value is noticeably higher for \emph{viral
nodes} ($\overline{r}_v = 2.96$) than for \emph{seed nodes}
($\overline{r}_s = 2.51$) showing a stronger involvement in viral
behavior when the invitation to pass messages along is received from
a trusted source. As a result, the average number of secondary cases
generated by each informed individual is given by the basic
reproductive number $R_0 = \lambda \overline{r_v}$. Both $\lambda$
and $\overline{r_v}$ also depend on the specific country in which
the campaign was run (see figure \ref{fig1b}) but in all cases we
found $R_{0}<1$, i.e. the viral campaigns did not reached the
``tipping-point''. Since the campaign execution was identical in all
countries, we conclude that differences observed in the propagation
parameters are due to the varying appeal of the viral offering to
customers in different markets. However, the data suggest a strong
linear correlation between the Transmissibility $\lambda$ and the
Fanout coefficient. This peculiarity of information diffusion
processes, not observed in traditional epidemics, stems from the
fact that the decisions of becoming a spreader and of the number of
viral messages to send, are taken by the same individual and thus
are, in average, correlated. As a result, the basic reproductive
number $R_{0}$ scales at least quadratically with the probability of
a touched individual becoming a {\em spreader}, i.e. being convinced
to propagate the message. Thus, increasing the perceived value of
the viral campaign offer would have a quadratic effect instead of a
linear one and the tipping-point would be reached for lower than
expected $\lambda$ values.

\medskip

\begin{figure}
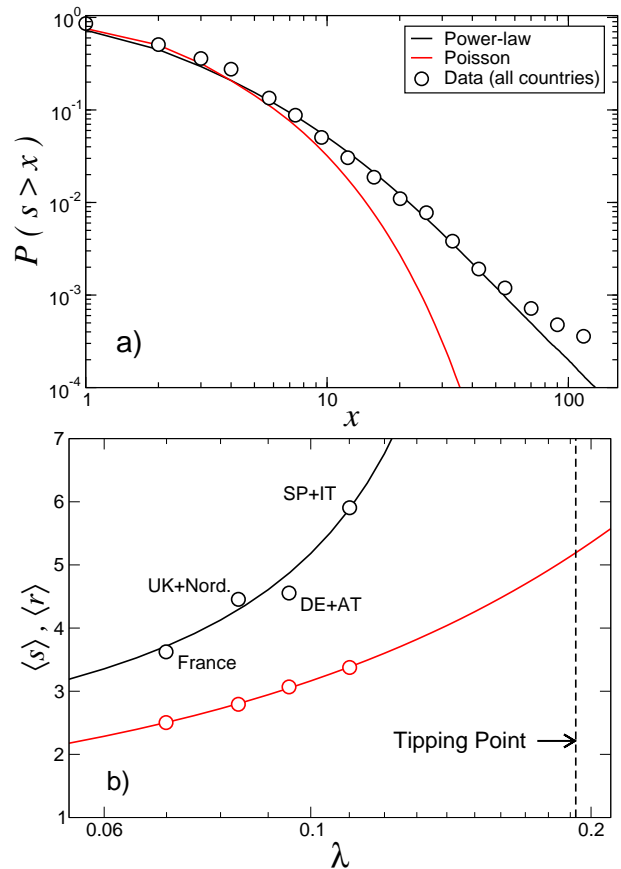

\centering
\includegraphics[width=0.45\textwidth,clip=]{ps.eps}\ \ \ \ \ \ \
\includegraphics[width=0.44\textwidth,clip=]{tippingpoint.eps}
\caption{{\bf a)} Cumulative distribution function of the
viral cascades size in all countries (circles). The solid black
line represents the prediction of the branching model (see text)
while the red solid line is the Poisson prediction. {\bf b)}
Average size of the viral cascades as a function of the Viral
Transmissibility $\lambda$ for different groups of countries
(circles). The solid line is the prediction of the branching model
(Eq. \ref{clustersize}) which diverges at the tipping point
$\lambda_c \simeq 0.1926$ estimated using the linear fits of
figure \ref{fig1b} for $\overline{r}_v$ and $\overline{r}_s$. The
red line and symbols shows $\overline{r}_v$ as a function of
$\lambda$. Note that at the tipping point the average number of
viral e-mails sent is just $\overline{r}_v = 5.18$.
 } \label{fig2}
\end{figure}

However, average quantities like $R_{0}$ can hide the heterogeneous
nature of information diffusion. In fact we find in our experiments
that most of the transmission we observe takes place due to
extraordinary events. In particular, we get that the number of
recommendations sent by \emph{spreaders} is distributed as a
power-law $P(r > x) \sim x^{-\alpha}$ as seen in figure \ref{fig1b},
indicating the high probability to find large number of
recommendations in the viral cascades. This large demographic
stochasticity has been observed in a number of other human
activities like the number of e-mails sent by individuals per day
\cite{barabasinature}, the number of telephone calls placed by users
\cite{telephone}, the number of weblogs posts by a single user 
\cite{blogs}, the number of web page clicks per user \cite{pitkow},
and the number of a person's social relationships \cite{tipping} or
sexual contacts \cite{sexual}. All these examples suggest that the
response of humans to a particular task cannot be described by
close-to-average models in which they behave in a similar fashion
probably with some small degree of demographic stochasticity. For
example we find that 2\% of the population has $r>10$, suggesting
the existence of super-spreading individuals in sharp contrast with
homogeneous models of information spreading \cite{bass}.
Super-spreading individuals have also been found in non-sexual
disease spreading \cite{ssdisease} where they have a profound
effect. As in that case, we find that super-spreading individuals
are responsible for making large viral cascades  rarer but more
explosive (see figure \ref{fig2}). For example, if we neglect the
existence of super-spreading individuals but still consider some
degree of stochasticity in the number of recommendations by making
$r$ a Poisson process with average $\overline{r}$, a viral cascade
like the one in figure \ref{fig1a} would have a probability of
appearance of approximately once every $10^{12}$ {\em seeds}, a
number much larger than the total world population (see figure
\ref{fig2}).

\medskip

An important question is whether the observed demographic
stochasticity in the number of recommendations is  directly related
to the heterogeneity of social contacts \cite{NM}. Recent available
data about social networks has revealed that humans show also large
variability in their number of social contacts. In particular, it
has been found that social connectivity is distributed as a
power-law, much like the number of recommendations in our viral
campaigns \cite{ebelemail}. Moreover, large variability in the
numbers of social contacts have a profound effect in information or
disease spreading \cite{epidemic,satorras}. Specifically,
simulations of information or disease spreading models on networks
show that if information or disease flows through {\em every} social
contact, the topological properties of social networks can
significantly lower the ``tipping-point''. While this might be the
case of computer virus spreading or any other kind of automatic
propagation through social networks, information transmission is
voluntary and participants who engage in the spreading consider the
cost and benefits of doing so. Thus, the number of recommendations
sent by each participant (including not sending any) results from a
trade-off between the information forwarding cost and the perceived
value of doing it. When the value is low, the average number of
recommendations can be very low, a small fraction of the sender's
social contacts which makes the social network topology largely
irrelevant in the decision making problem. In fact, our data suggest
that this is the case; specifically, most of the viral cascades have
a tree-like structure while social networks are characterized by the
large density of local loops \cite{why}. To illustrate this
observation quantitatively, we have measured the clustering
coefficient $C$, i.e., the fraction of an individual contacts who
are in contact between themselves. Email social networks have large
values of clustering ($C_{email} \sim 0.15 - 0.25$) \cite{NM} while
in our case we find $C_{viral} = 4.81\times 10^{-3}$. Of course,
these numbers are not independent: as shown in the appendix \ref{appendixc} and under fairly general assumptions we should expect that
$C_{viral} = C_{email}\times 2 R_{0}/(\langle\overline{k}_{nn}\rangle-1)$ where
$\overline{k}_{nn}$ is the average number of social contacts of the
neighbors of an individual. In social networks $\overline{k}_{nn}$
is a large number, and then viral cascades have a very small
clustering coefficient {\em even when close to the tipping-point}
$R_{0} \simeq 1$. Thus, we have found that reach of information
diffusion can be very large without sampling the topological
properties of the social network of individuals. This implies that
the large heterogeneity observed in the number of recommendations is
a characteristic of human decision making tasks rather than a
reflection of the social network.

\medskip

Given the above results, we have modeled the viral campaigns
recommendation cascades through a branching process in which the
recommendation heterogeneity is considered but the social network
topology is neglected. Each cascade starts from an initial {\em
seed} that initiates viral propagation with a random number of
recommendations distributed by $P(r_{s})$ and whose average is
$\overline{r}_{s}$. Touched individuals become secondary spreaders
with probability $\lambda$ thereby giving birth to a new generation
of viral nodes which, in turn, propagate the message further with
$r_{v}$ recommendations distributed by $P(r_{v})$ with average
$\overline{r}_{v}$ \footnote{Actually, the distributions $P(r_{s})$
and $P(r_{v})$ are different but we use the same letter for clarity.
See appendix \ref{appendixa} for more information}. The propagation continues
through successive generations until none of the last touched
individuals decide to become secondary spreaders. This process
corresponds to the well known Bellman-Harris branching model
\cite{branching}. On average, the infinite time limit cascade size
can be estimated as
\begin{equation}\label{clustersize}
\overline{s} = 1 + \frac{\overline{r}_{s}}{1-R_{0}}
\end{equation}
which are within a striking $1\%$ error of the experimental values
found in the viral campaigns (see Table \ref{table1}). Not only are average cascade sizes
well predicted, but their distribution is properly replicated when
the heterogeneity in the number of recommendations is implemented
(see figure \ref{fig2}). Both results show how accurate the model
can be in predicting the extent of a viral marketing campaign: since
the values of $\lambda$ and $\overline{r}_{v},\overline{r}_{s}$ can
be roughly estimated during the early stages of the campaign, we
could have predicted the final reach of a viral campaign at its very
beginning. Moreover, giving the knowledge of how $\lambda$ and
$\overline{r}_{v}$ are connected and using equation
(\ref{clustersize}) we could give estimations of the critical viral
transmissibility $\lambda_{c}$ which makes the viral message
percolate through a fraction of the entire network \footnote{Since
e-mail Networks carrying viral propagation are semidirected
\cite{NM} some portions of them are unreachable due to lack of
connecting paths. So, we define percolation as the state where
messages reach a large fraction of the e-mail Network Giant
Connected Component (GCC)}. We found that $\lambda_{c}=0.1926$ which
correspond to $\overline{r}_{v} = 5.18$. Of course this is an upper
limit to the real ``tipping-point'' since it is based on the
assumption that each {\em seed} originates one isolated viral
cascade, which is only valid far from the ``tipping-point''. The low
number of recommendations needed to reach the ``tipping point''
illustrates the limited effect of the social network topology in the
efficiency of viral campaigns. Thus, it is not necessary to send the
message to each participants' social contact in order to reach a
significant fraction of the target population.

\medskip

\begin{figure}
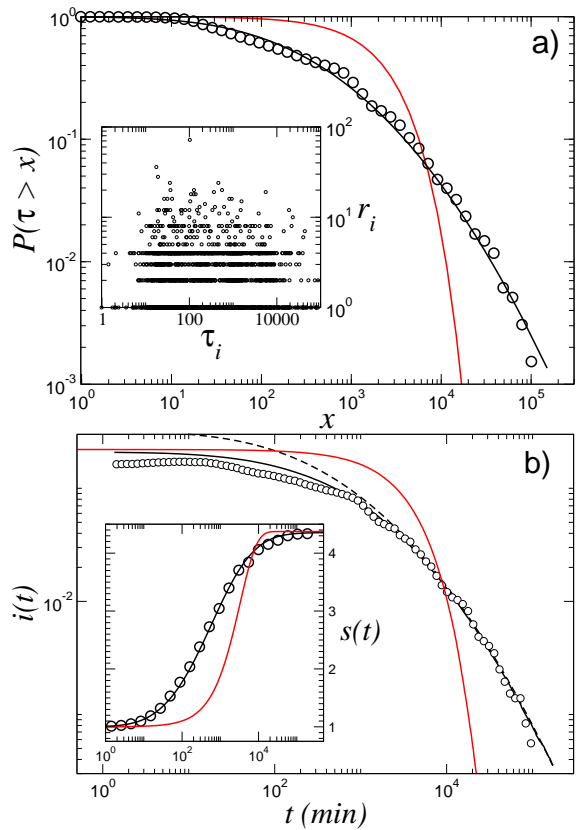

\centering
\includegraphics[width=7.5cm,clip=]{tau.eps}\\
\includegraphics[width=7.5cm,clip=]{tau1.eps}
\caption{\textbf{a)} Cumulative probability distribution of time elapsed$\tau$ between the reception and forwarding of the viral information
(circles) for participants in all countries. The solid line shows
MLE fit to a log-normal distribution with $\hat \mu = 5.547$ and
$\hat \sigma^2 = 4.519$. Only viral nodes are considered, since
reception time for seed nodes is undefined. Inset shows absence of
statistical correlation between the number of recommendations made
$r_{i}$ and the time elapsed $\tau_{i}$ until each participant
forwards the message. \textbf{b)} Average number of touched
participants as a function of the cascades start time in our
campaigns (circles) compared with the prediction of the
Bellman-Harris model (solid line), with the fitted log-normal
distribution (black), and with an exponential distribution of the
same mean (red). The dashed line is the analytical approximation to
a Bellman-Harris process with log-normal waiting times given by
$i(t) = 1/(1-\lambda \overline{r}_v) [1-G(t)]$, where $G(t)$ is the
cumulative distribution function of the log-normal distribution in
a). Inset: Remarkable agreement between the average size of the
viral cascades as function of total campaign time in log scale
(circles) with the Bellman-Harris model prediction with \emph{G(t)}
log-normal. Also shown, in red, the prediction with \emph{G(t)}
exponential.}\label{fig3}
\end{figure}

Information diffusion dynamics is also affected by the different way
individuals program the execution of their tasks. The time it takes
for participants to pass the message along since it was received, or
``waiting-time'' $\tau$, shows also a large degree of variability:
participants forward the message after $\overline{\tau}=1.5$ days on
average, but with a very large standard deviation of
$\sigma_{\tau}=5.5$ days, with some participants responding as late
as $\tau = 69$ days after receiving the invitation email (see figure \ref{fig3}). The large
variability of the distribution $G(\tau)$ for waiting times observed
in our data is consistent with recent measures of how humans
organize their time when working on specific tasks, such as email
answering, market trading or web pages visits.
\cite{barabasinature,vazquez}. Traditional Poissonian models for
$G(\tau)$ cannot match the observed data and several long-tailed
models like power laws \cite{vazquez} or log-normal
\cite{amaralemail} distributions for $G(\tau)$ have been proposed to
incorporate the large waiting-times between actions observed. Our
data is fully consistent with a log-normal distribution and,
moreover, the data shows no statistical correlation with the number
of recommendations made by the participant (see figure \ref{fig3}).
This
means that the delay in passing along a message and the number of
recommendations made by individuals are largely independent
decisions. Within this approximation, our simulations of the
Bellman-Harris process with waiting times distributed by log-normal
$G(\tau)$ and number of recommendations by the power-law $P(r)$ show
a remarkable agreement with our data from the campaigns (see figure
\ref{fig3}). On the other hand, population-average models predict that the
average number of infected individuals $i(t)$ passing along the
message at time $t$ is described by the growth equation
\begin{equation}\label{malthusian}
\frac{d i}{dt}  = \alpha_{0} i
\end{equation}
where $\alpha_{0} = (R_{0}-1)/\overline{\tau}$ is the Malthusian
rate parameter of the population. The number of people aware of the
information until time $t$ is the cumulative sum of infected
individuals, $s(t) = \int_{0}^t i(s) ds$. Equation
(\ref{malthusian}) is the starting point of many different
deterministic models to describe the evolution of epidemics,
information or innovations in a population. It also describes the
asymptotic dynamics of those situations in the models with some mild
degree of heterogeneity in $\tau$ \footnote{If $G(\tau)$ is
Poissonian, the average number of infected people in Bellman-Harris
process is given {\em exactly} by equation (\ref{malthusian})}. The
situation changes drastically when $G(\tau)$ has a large degree of
variability. Specifically, if $G(\tau)$ belongs to the so-called
class of {\em subexponential distributions}, i.e. distributions that
decay slower than exponentially when $\tau \to \infty$, equation
(\ref{malthusian}) is not valid. This class contains important
instances as power-law (or Pareto) distribution, the Weibull or,
like in our case, the log-normal distribution. In the latter we
obtain that for $R_{0}<1$, $i(t)$ is given in the long run by
\begin{equation}\label{malthusian1}
i(t) \sim \frac{1}{1-R_{0}}[1-\int_{0}^t G(\tau)d\tau] \sim \frac{1}{1-R_{0}}e^{-a\ln^2 t}/\ln t
\end{equation}
with $a>0$ a constant independent of $R_{0}$ (see appendix \ref{appendixb}). Equation (\ref{malthusian1}) demonstrates the deep impact
of large degree of heterogeneity in our population: the very
functional form of the time dependence is changed and the dynamics
of the system depends on a logarithmic time scale, thus slowing down
the propagation of information in a drastic way. The situation is
the opposite for moderate values of $R_{0} > 1$ where $i(t) \sim
e^{\alpha t}$ with $\alpha$ given by the solutions of
$R_{0}\int_{0}^\infty e^{-\alpha t}G(t)dt = 1$ but with $\alpha \gg
\alpha_{0}$ and thus information spreads much faster than expected.
The different behavior both above and below the ``tipping-point'' is
due to the different importance that individuals with small or large
values of $\tau$ have in the dynamics: while below $R_{0}=1$ the
number of infected individuals decay in time  up to the point where
a sole individual can halt the dynamics of a viral cascade, above
$R_{0} > 1$ the dynamics is governed by individuals with small
number of $\tau$ which are more abundant than those with $\tau
\simeq \overline{\tau}$ and thus speed up the diffusion. Since
subexponential distributions are found in other human tasks
\cite{barabasinature,vazquez,amaralemail}, our findings have the
important consequence that the high variability in the response of
humans to a particular task can slow down or speed up the dynamics
of processes taking place on social networks when compared to the
traditional population-average models.

\begin{figure}
\centering
\includegraphics[width=0.45\textwidth,clip=]{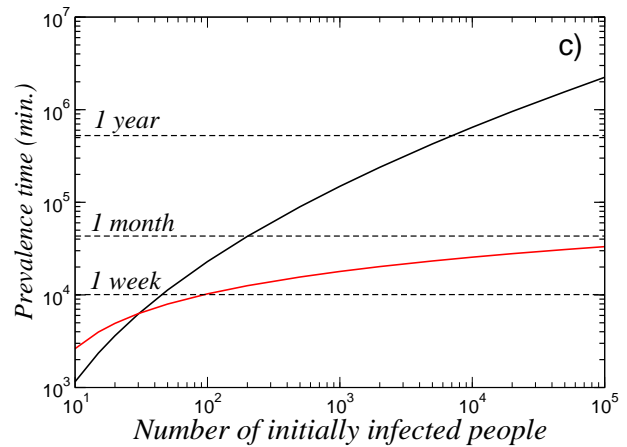}
\caption{ Prevalence time $t_{f}$ as a function of
number of initially infected people (i.e. number of seeds $N_{s}$)
for the Bellman-Harris branching process with values of $R_{0}
=\lambda \overline{r}_{v}$ and $\overline{r}_{s}$ obtained in our
campaigns for all countries (see table \ref{table1}). Prevalence
time is calculated by solving equation $i(t_{f}) = 1/N_{s}$. Solid
lines correspond to different distributions $G(\tau)$: log-normal
(black) and Poisson (red).
 } \label{fig4}
\end{figure}

\medskip

Our study does not explain why the frequency and number of
recommendations made by people in our experiments are so
heterogeneous despite the decision they faced was the same. Rational
expectations suggest that individuals should have made their
decisions based on similar utility functions and then the answers
would have been closer to each other. The fact that the same degree
of heterogeneity has been found for so many different tasks in
humans \cite{barabasinature,vazquez,amaralemail} suggest that it is
an intrinsic feature of human nature to be so wildly heterogeneous.
As we have shown, the main consequence of the large variability of
human behavior is that population-level average quantities do not
explain the dynamics of social network processes. Important
consequences of this large variability of behavior are the slowing
down or speed up of information diffusion and that most of the
diffusion takes place due to otherwise considered extraordinary
events. The corrections to population-averaged predictions go beyond
a different set of values for the dynamics parameters: They can even
change the time scale or functional form of the predictions. In
particular, we have seen that we are forced to revisit the way we
model spreading processes mediated by humans by using differential
equations like (\ref{malthusian}). On the other hand, the slowing
down of information diffusion implies that viral cascades or
outbreaks do last much longer than expected, which could explain the
prevalence of some informations, rumors or computer viruses. For
example, if we assume that initially $N_{s}$ seeds are infected, we
could take as the end of information diffusion the point when the
fraction of infected individuals decays to $i(t_{f}) \sim 1/N_{s}$.
While Poissonian approximations yield to $t_{f} \simeq
\overline{\tau}/(1-R_{0})\ln N_{s}$, in our case we find that $t_{f}
\sim e^{\sqrt{b \ln N_{s}}}$ where $b>0$ is independent of $R_{0}$.
When $N_{s}$ is large enough there is a huge difference between both
estimations. For example, if $N_{s}=10^4$ (a large but moderate
value), then $t_{f} = 17$ days (with $R_{0} = \lambda
\overline{r}_{v}$) for Poissonian models while $t_{f} \simeq 1$ year
if $G(\tau)$ is described by a log-normal distribution. As suggested
in \cite{barabasivirus}, the high variability of response times can
be the origin of the prevalence of computer viruses. In fact, our
viral cascades span in time longer than initially expected, which
may render viral campaigns unpractical for information diffusion.
Companies, organizations or individuals implementing such marketing
tactics to disseminate information over social networks face the
following dichotomy: If the tactic is successful and information
spread reaches the ``tipping-point'' it does so very quickly;
however, if it fails in reaching the ``tipping-point'', the
situation is even worse because information travels slowly in
logarithmic time. We hope that our experiments and the fact that
they can be accurately explained by simple models will trigger more
research to understand quantitatively human behavior.

\medskip

{\bf Acknowledgments:} J.L.I. acknowledges IBM Corporation support
for the collection of anonymous data of its Viral Marketing
campaigns propagation. E.M. acknowledges partial support from MEC
(Spain) through grant FIS2004-01001 and a Ram\'on y Cajal
contract. We thank Alex Arenas for sharing with us the e-mail Network data used in our simulations.
\medskip

\appendix

\section{Model Selection}\label{appendixa}
\subsection{Candidate Models for the recommendation distribution}

The recommendation distribution is the probability distribution of
the number of recommendations $r$ made by each participant in the
campaign. As shown in figure 1b, there is a large degree of
heterogeneity in the way the participants engaged in the campaign.
The number of recommendations per participant varies from one to
more than one hundred and thus any modeling of the distribution of
recommendations has to incorporate those extreme events.

We consider two distinct treatments of the number of
recommendations:
\begin{enumerate}
\item In order to incorporate demographic stochasticity inherent
to the transmission process, many classical epidemiological models
assume that the offspring distribution is represented by a Poisson
process, and thus $r \sim Poisson(\mean{r})$.
\item However, there is an increasing evidence that humans tend to
respond in a untamed way in different activities. Most people
behave close to the average behavior, but a not negligible portion
of humans show bursts of activities, like the number of e-mails
sent per day \cite{ebelemail}, the number of telephone calls
placed by users \cite{telephone}, the number of weblogs posts by a
single user\cite{blogs}, the time spent between receiving and
replying an e-mail \cite{barabasinature} or the number of web page
clicks per user \cite{pitkow}. To account for those extreme
events, power-law distributions of activity have been proposed and
observed statistically. Here we propose a model for the number of
recommendations based on a power-law distribution $r \sim
\mathrm{PL}(\alpha,\beta)$ which has the following pdf
\begin{equation}\label{PL}
P_{\mathrm{PL}}(r) = \frac{H_{\alpha,\beta}}{\beta + r^\alpha}
\end{equation}
which asymptotically decreases like a power law and shows a cutoff
at small numbers of recommendations $r^* \simeq \beta^{1/\alpha}$.
Here, $H_{\alpha,\beta}$ is a normalization constant so that
$\sum_{r=1}^\infty P(r) = 1$.
\end{enumerate}

\subsection{Parameter estimation}
We estimate the model parameters by the method of moments to ensure
that all models have the same mean value $\mean{r}$ (and $R_{0}$)
observed in the campaigns, so that the difference between models is
due to the different way they handle heterogeneity. Note that the
Poisson distribution has only one parameter and then only $\mean{r}$
can be fitted. In the other case, the $\mathrm{PL}(\alpha,\beta)$,
there are two parameters and data can be fitted to the first and
second moment of $r$ as shown in table \ref{suptable}. We model
independently the pdf of the number of recommendations made by seeds
and viral nodes to account for the different $\overline{r}$ values
observed. It is interesting to note that both pdfs seem to decay as
a power law with the same exponent $\alpha \simeq 3.5$.

\begin{table}
\centering
\begin{tabular}{c|cc|cc}
\hline
Group & $\overline{r}$ & $\overline{r^2}$ & $\alpha$ & $\beta$ \\ \hline
Seeds & $2.51$ & $15.2$ & 3.48 & 29.66\\
Viral & $2.96$ & $20.5$ & 3.50 & 60.07\\\hline
\end{tabular}
\caption{\label{suptable}Parameters of the different probability
distribution models for the observed number of recommendations made
by seed nodes and viral nodes. Parameters $\alpha$ and $\beta$ refer
to }
\end{table}

\section{Viral Marketing propagation dynamics}\label{appendixb}

\subsection{The Galton-Watson branching process}
Branching processes describe the evolution of systems where an
initial set of objects called the 0-th generation reproduce
themselves into a set of children of the same kind call the first
generation and so on through successive generations. The
Galton-Watson process is the simplest mathematical description of
such situation and only keeps track of the sizes of the successive
generations, not the times at which individual objects are born or
their individual family relationships. We can define two sets of
random variables $\{G_n\}=\{G_0,G_1,G_2, ....\}$ with $G_n$ being
the number of individuals in generation $n$ and
$\{F_n\}=\{F_0,F_1,F_2, ...\}$ with $F_n=\sum_{i=0}^nG_i$. Since
the probability law governing each generation does not depend on
the sizes of the preceding generation, both form a \emph{Markov
Chain}.

\medskip

The probability distribution of the variable $G_1$ is given by $
P(G_1=k)=p_k$ and we can define its probability generating function (pgf) $f(s)$ as
\begin{equation}\label{f(s)}
    f(s)=\sum_{n=0}^\infty p_ns^n
\end{equation}
whose derivative evaluated at $s=1$ is the expected value of $G_1$
as follows
\begin{equation}\label{average}
    \mean{G_1}\equiv m=f'(1)=\sum_{n=0}^\infty np_n
\end{equation}
It was demonstrated by Watson \cite{harris} that the generating
function of $G_n$ is $f_n(s)$, the n-th iterate of the generating
function $f(s)$, as follows
\begin{equation}\label{iterate}
    f_n(s)=f\{f[...f(s)...]\}
\end{equation}
This important property leads to the following result for the
average size of the n-th generation:
\begin{equation}\label{averagen}
    \mean{G_n}=f'_n(1)=(f'(1))^n=m^n
\end{equation}

\subsection{Model for Viral Marketing propagation}
Applying the Galton-Watson formalism to the viral propagation
dynamics, we consider a single propagation tree starting from one
node ($G_0=1$) whose components are all nodes touched by the
message. Its total size at generation $n$ is $F_n=\sum_{i=0}^nG_i$
and the nodes can be divided in \emph{Active} ($F_n^A$) and
Passive ($F_n^P=F_n-F_n^A$) depending on whether they have passed
the viral message along or not. Now, we define the \emph{Viral
Transmissibility}, or the probability of any one node being
Active, as $\lambda=F_n^A/F_n$ and the \emph{Fanout Coefficient},
or average number of email referrals sent by Active nodes, as
$\overline{r}_v=[\sum_{n=1}^{F_n^A}r_n]/F_n^A$ where $r_n$ is the
number of email referrals sent by node n. Now the average number
of email referrals sent by all nodes (Active or Passive) is
\begin{eqnarray}\label{allfanout}
    \sum_{r=0}^{F_n}rp_r &=&\frac{1}{F_n}\sum_{n=1}^{F_n}r_n=\frac{1}{F_n}\left[\sum_{n=1}^{F_n^A}r_n-\sum_{n=F_n^A+1}^{F_n}r_n\right]\\
    \nonumber
    &=&\frac{F_n^A}{F_n}\texttt{ }\overline{r}_v=\lambda\overline{r}_v
\end{eqnarray}
since summation over Inactive nodes is zero. In our mean-field
approach, this value will be considered to be constant throught
all generations.

\medskip

Now, the probability function of the Galton-Watson process is
given by $p_0=1-\lambda$, $p_r\{1,2,....\}$ where $p_r$ is the
power-law distribution in (\ref{PL}) with
$\sum_{r=0}^{\infty}p_r=1$, $\sum_{r=1}^{\infty}p_r=\lambda$ and
$\sum_{r=0}^{\infty}rp_r=\lambda\overline{r}_v$. The corresponding
generating function is
\begin{equation}\label{viralgen}
    f(s)=1-\lambda+\sum^{\infty}_{r=1}p_rs^{r}
\end{equation}
and applying the Galton-Watson process results in (\ref{average})
and (\ref{averagen}) we write the average size of each of the
generations in the propagation tree as
\begin{equation}\label{G1}
    \mean{G_1}\equiv R_0=f'(1)=\sum_{r=0}^{\infty}rp_r=\lambda\overline{r}_v
\end{equation}
and
\begin{equation}\label{Gn}
    \mean{G_n}=f'_n(1)=[f'(1)]^n=R_0^n=(\lambda\overline{r}_v)^n
\end{equation}
hence, the average size of a branch in the mean-field approach at
the infinite time limit is given by
\begin{equation}\label{Fn}
    F_\infty=\mean{\sum_{n=0}^{\infty}G_n}=\sum_{n=0}^{\infty}\mean{G_n}=\sum_{n=0}^{\infty}(\lambda\overline{r}_v)^n=\frac{1}{1-\lambda\overline{r}_v}
\end{equation}
since the summation converges because the system is below the
percolation threshold and $\lambda\overline{r}_v<1$. Now, the
total number $N$ of nodes in the Viral Network graph in the
infinite time limit results from adding the nodes in the
$\overline{r}_s$ trees generated by each \emph{seed node} and
multiplying by the total number $N_s$ of \emph{seed nodes}. Thus
we have, \emph{seed nodes} included, that
\begin{equation}\label{N}
    N=N_s+N_s\overline{r}_vF_\infty=N_s
    \left(1+\frac{\overline{r}_s}{1-\lambda\overline{r}_v}\right)
\end{equation}
where the validity condition of being far from the percolation
threshold is necessary to ensure that outbreaks (or clusters)
originating from different \emph{seed nodes} do not merge with one
another.

\subsection{Age-dependent dynamics: Bellman-Harris process}

The description of viral marketing dynamics based on the
Galton-Watson process does not consider the "waiting time"
($\tau$) elapsed between the reception of a message and the moment
its passing along, assuming implicitly that both actions take
place at the same instant. However, viral propagation does not
occur instantaneously and our experiments show that it follows a
log-normal time distribution much like those observed in other
human activities.

\medskip

To describe this behavior we will use the Bellman-Harris process, a
continuous time generalization of the Galton-Watson one, in which
both the number of descendants at each generation and their
lifetimes are represented by non-negative, independent random
variables \cite{harris}. It is described as follows: A single
ancestor is originated at $t=0$ and lives for time $\tau$ which is a
random variable with cumulative distribution function ${G(\tau)}$
with mean $\overline{\tau}$. At the moment of its disappearance the
particle generates a number $r$ of progeny according to a
probability distribution $P(r)$ whose pgf is denoted as $f(s)$. The
process continues with descendants behaving independently and in the
same fashion as their ancestors did. Thus, the branching process is
described by the random variable $Z(t)$ representing the number of
active particles at time $t$. In our case, $Z(t)$ represents the
number of active participants at time $t$, i.e.\ the number of
people that have received the information before time $t$ and that
will send it in a future time.

\medskip

Analytically, we use the generating function $F(s,t)$ for
calculating the probability of having $Z(t)$ particles
active at time $t$. It is defined as
\begin{equation}\label{bh1}
    F(s,t)=\sum^{\infty}_{i=0}P(Z(t)=i)\;s^{i}
\end{equation}
It can be proved \cite{harris} that $F(s,t)$ in the
asymptotic limit satisfies a renewal equation of the form
\begin{equation}\label{bh2}
    F(s,t)=s[1-G(t)]+\int^{\infty}_{0}dG(\tau)\;f[F(s,t-\tau)]
\end{equation}
As a result $i(t)$, the expected value of $Z(t)$, verifies
that
\begin{equation}\label{bh3}
    i(t)=\frac{\partial{F}}{\partial{s}}(1,t)=1-G(t)+R_0\int^{t}_{0}dG(\tau)\;\emph{i}(t-\tau)
\end{equation}
where we have used that
\begin{equation}\label{deriv}
    \left.\frac{\partial{f[F(s,t-\tau)]}}{\partial{s}}\right|_{s=1}=\left.\frac{\partial{f(s)}}{\partial{s}}\right|_{s=1}\left.\frac{\partial{F(s,t-\tau)}}{\partial{s}}\right|_{s=1}=R_0\;i(t-\tau)
\end{equation}
General explicit solutions of the integral equation (\ref{bh3}) do
not exist, although the asymptotic behavior is known in the case in
which the Malthusian parameter $\alpha$ of the population exists.
This parameter is defined explicitly by
\begin{equation}\label{malthusianapp}
R_0 \int_0^\infty e^{-\alpha t} d G(t) = 1.
\end{equation}If a solution of this equation exists, then \cite{harris}
\begin{equation}
i(t) \sim C e^{\alpha t},\qquad C=\frac{R_0-1}{\alpha R_0^2\int_0^\infty t e^{-\alpha t} dG(t)}
\end{equation}
The normalization of $G(t)$ implies that, if exists, $\alpha > 0$
for $R_0 > 1$ and $\alpha <0$ for $R_0 < 1$ thus recovering the
exponential growth or decay above and below the ``tipping-point".
Important instances of this case are:
\begin{enumerate}
\item \textbf{Galton-Watson process.} For $G(t)=\chi(t-\overline{\tau})$,
where ${\chi(t)}$ is the unit step function at 0 (i.e., lifespan
of all particles is identical and equal to $\tau$), we recover a
Galton-Watson process with progeny generating function
$f(s)$ and mean
\begin{equation}\label{bh4}
    i(t=n\overline{\tau})=R_0^{t/\overline{\tau}}
\end{equation}
which yields to equation (\ref{N}) since
$R_0=\lambda\overline{r}_{v}$. \item \textbf{Markov age-dependent
branching process.} Traditional modeling of the lifespan or
``waiting time" of human activities implies that $G(t)$ is of the
Poissonian type $G(t)=1-e^{-t/\overline{\tau}}$. One of the
important reasons is that this exponential distribution has the {\em
lack-of-memory property} which is suitable for modeling the dynamics
using Markovian processes. This is exemplified in our case by the
fact that, if $G(t)$ is exponentially distributed, then the solution
of (\ref{bh3}) is {\em exactly} given by
\begin{equation}
i(t)=e^{\alpha_0 t},\qquad \alpha_0 = \frac{R_0-1}{\overline{\tau}}
\end{equation}
\end{enumerate}
Note that both cases correspond to the basic Markovian growth models
of epidemic transmission in which the average number of infected
people grows or decays exponentially within a time scale
proportional to the average lifespan of infected individuals.

\medskip

However, the Malthusian parameter of the population does not exist
when $R_0<1$ for a broad and important class of distributions called
{\em sub-exponential distributions}: a probability distribution with
cdf $G(t)$ defined on $[0,\infty)$ is said to be subexponential if
$\overline{G^{*2}}(t)\sim{2\overline{G}(t)}$ as $t\rightarrow\infty$
where $\overline{G}(t)=1-G(t)$ and $G^{*n}$ denotes the n-fold
convolution of function $G(t)$ by itself. As a consequence of this
asymptotic behavior, the integral in  (\ref{malthusianapp}) does not
exist for $\alpha < 0$ which means that the pdf of this class of
distributions decays slower than any exponential when $t\to\infty$.
Important instances like the Pareto, log-normal and Weibull
distributions belong to this category. In this case, the solution of
(\ref{bh3}) is a non-Markovian and the usual modeling of epidemics
in terms of growth equations or differential equations fails: in
particular, the knowledge of how information has been diffused until
time $t$ does not determine the dynamics for longer times. The
general asymptotic behavior of equation (\ref{bh3}) is known to be
of the form \cite{athreya}
\begin{equation}\label{bh5}
    i(t)\sim\frac{1}{1-R_0}\;\overline{G}(t),
\end{equation}
and thus the number of infected people decays like the tail of the distribution.

\medskip

We have analyzed the evolution of viral campaigns and found that the
average cascade size as a function of time
$s(t)=\int^{\infty}_{0}i(\tau)d\tau$ can be modeled with remarkable
precision by a Bellman-Harris process as in (\ref{bh5}) with $G(t)$
lognormal. Thus, instead of observing the usual exponential decay of
active people $i(t)\sim{e^{{\alpha}t}}$ the active viral population
evolves as
\begin{eqnarray}
  i(t)&\sim&\frac{1}{2(1-R_0)} \left[\mathrm{Erf}\left(\frac{\mu - \ln t}{\sqrt{2\sigma^2}}\right)-1\right]\\ &\sim&\frac{1}{(1-R_0)} \frac{\sigma}{\sqrt{2\pi}} \frac{\exp\left(-\frac{(\mu-\ln t)^2}{2\sigma^2}\right)}{\ln t - \mu}
\end{eqnarray}
for large $t$. The asymptotic behavior depends then on a different
time scale (logarithmic in time $\ln t$) rather than the normal time
scale $t$, a result that highlights the failure of typical modeling
to explain observed behavior when the variability of humans is so
large than it is described by a subexponential distribution.

Note that the influence of the log-normal distributions of waiting
times occurs even at the population average level and not only on
fluctuations around the average value $i(t)$, i.e., it changes the
dynamics not just quantitatively but also qualitatively. Finally,
the dynamics is slowed down by the high probability of finding an
individual with large response times, as the logarithmic time scale
in our case shows.

\begin{figure}
\centering
\includegraphics[width=8cm,clip=]{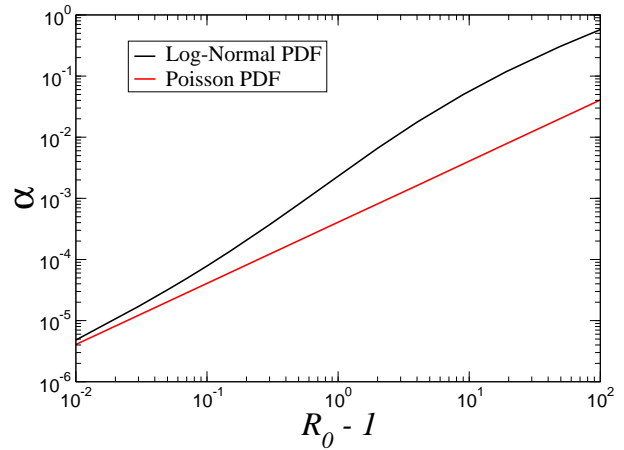}
\caption{ Malthusian parameter of the population above the
``tipping-point" as a function of the average number of secondary
cases for different distributions of $G(t)$.}\label{figmalth}
\end{figure}

\medskip

For $R_0>1$ the Malthusian parameter exists for the class of
subexponential distributions and then $i(t)$ grows exponentially
like $i(t) \sim e^{\alpha t}$. But, even in this case, there is a
large quantitative difference between the solutions of equation
(\ref{malthusianapp}) and the values expected by assuming exponential
distributions. As shown in figure \ref{figmalth} the difference in
our case can be of one order of magnitude which implies that if the
campaign reaches the tipping-point the information spreads much
faster than expected. For example, if $R_0 = 2$ and using the values
of $\overline{\tau} \simeq 1.5$ days obtained in our campaigns we
should have expected an exponential growth with time scale
$\alpha_0^{-1} = \overline{\tau} \simeq 1.5$ days, while in the case
of a log-normal distribution we get $\alpha_0^{-1} \simeq 7$ hours.
This large quantitative difference is due to the fact that
subexponential distributions are more skewed than the Poisson ones
and thus there is a higher probability of finding participants with
small ``waiting-times'' (compared to the mean) in subexponential
distributions. Those fast responders are responsible for this
exponential growth with shorter time scale.

\section{Inferences on the substrate e-mail Network}
\label{appendixc}
The e-mail Network serving as substrate of the viral messages
propagation is formed by individuals (nodes) and by their e-mail
connections (links between nodes) as determined by the addresses
listed in their e-mail address books. In their propagation, viral
messages can only go through the links in the e-mail Network and
the viral network is thus a subset of it. We have observed
however, that even when viral propagation has fully percolated,
the substrate e-mail Network is not readily perceived through
observation of the Viral Network.

\medskip

\begin{figure}
\centering
\includegraphics[width=8cm,clip=]{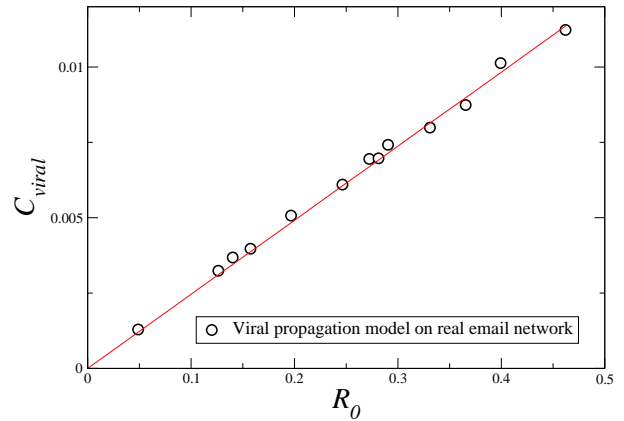}
\caption{Clustering coefficient $C_{viral}$ for the viral cascades
obtained through simulations of the viral propagation model on a
real email network (symbols) compared with the lineal relationship
given by equation (\ref{eq8}). The email network has $C_{email} =
0.2202$ and $\mean{\overline{k}_{nn}} = 18.903.$}\label{figclust}
\end{figure}

Nevertheless, because both networks are related, some parameters in
the e-mail Network can be gleaned through measures on the viral
network. We prove here that in a viral propagation process the
clustering coefficients of the substrate network (the e-mail
Network) and of its virally percolated subset (the Viral Network)
are correlated and derive, based on a mean-field approximation, an
expression of such correlation. The clustering coefficient,
according to Watts and Strogatz \cite{clustering}, is defined as
\begin{equation}
\label{eq1} C=\frac{1}{N}\sum\limits_i^N
{\frac{number\;of\;triangles\;connected\;to\;node\;i}{number\;of\;triples\;centered\;on\;node\;i}}
\end{equation}
where "triple" means a single node with edges running to an
unordered pair of others. If such pair is also connected, it forms
a triangle or "transitive triad". Now we can write, in a
mean-field approximation, the clustering coefficients of the
e-Mail and Viral networks respectively as
\begin{equation}
\label{eq2} C_{email} =\frac{1}{N_e }\sum\limits_i^{N_e }
{\frac{(triang_{email})_i }{(triples_{email})_i }\sim
\frac{\langle (triang_{email})_i\rangle }{\langle
(triples_{email})_i\rangle }}
\end{equation}
\begin{equation}
\label{eq3} C_{viral} =\frac{1}{N_v }\sum\limits_i^{Nv}
{\frac{(triang_{viral})_i }{(triples_{viral})_i }}\sim
\frac{\langle(triang_{viral})_i
\rangle}{\langle(triples_{viral})_i\rangle }
\end{equation}
Considering an e-mail Network node connected to triangles and
triples, we can watch the bond percolation progress of a viral
message planted on it. The probability of a triangle on such node
being fully percolated by e-mails is the joint probability of
percolation of each of the edges in the triple and of the link
between the two neighbors at the end of them which forms the
triangle third side
\begin{equation}\label{eq4}
P(perc\_ triang.)=P(perc\_ triple)\;\times \;P(perc\_ 3rd\_ side)
\end{equation}
As a result, we can estimate as follows the average number of
triangles and triples in the Viral Network with the mean-field
approximation
\begin{eqnarray}
\langle(triang_{viral})_i\rangle&=&P(perc\_triple)\times
P(perc\_3rd\_side)\times \nonumber\\ & & \times \langle(triang_{email})_i\rangle \label{eq9}
\end{eqnarray}
\begin{equation}\label{eq5}
\langle(triples_{viral})_i\rangle =P(perc\_triple)\times\langle
(triples_{email})_i\rangle
\end{equation}
Combining (\ref{eq2}), (\ref{eq3}), (\ref{eq9}) and (\ref{eq5}) we
obtain
\begin{equation}\label{eq6}
C_{viral}\simeq P(perc\_3rd\_side)\times\frac{\langle(triang_{email})_i\rangle}{\langle
(triples_{email})_i\rangle}
\end{equation}
Considering that the clustering coefficient is calculated for
non-directed networks (i.e. arcs in the e-mail Network are
assimilated to undirected edges), that nodes reached by the viral
message become active with probability \textit{$\lambda $} (the
Transmissibility) and that, after becoming active they send
messages with Fanout $\overline{r}_v $ each, we conclude that the
probability for the third side of the triple being percolated by a
viral message, so as to close a triangle, is given by
\begin{equation}\label{eq7}
P(perc\_3rd\_side)=\frac{2\lambda\overline{r}_v}{\mean{\overline{k}_{nn}}_e-1}
=\frac{2R_0}{\mean{\overline{k}_{nn}}_e-1}
\end{equation}
where $\mean{\overline{k}_{nn}}_e$ is the average over the email
network of the nearest neighbors average degree. It has to be
decreased by 1 because the propagation rules do not allow messages
to be sent back to ancestor nodes. The factor 2 results from the
fact that either of the two nodes at the open end of a triple can
send the message that closes the corresponding triangle.
Substituting (\ref{eq7}) and (\ref{eq2}) in (\ref{eq6}) we arrive to
the relationship between an e-mail Network clustering coefficient
and that of its virally percolated one
\begin{equation}\label{eq8}
C_{viral} \simeq \frac{2R_0}{\mean{\overline{k}_{nn}}_e-1}\times
C_{email}
\end{equation}
This expression has been tested through simulations of the viral
propagation model on a real email network gathered from email server
logs of a Spanish university \cite{alex} (see figure
\ref{figclust}). In the model, any node becomes a secondary spreader
with probability $\lambda$ and transmits the message among $r$ of
his/her email connections (if possible) with average
$\overline{r}_v$ number of recommendations. While the real network
has a rather large clustering coefficient $C_{email} \simeq 0.22$,
the resulting viral cascades have a very small clustering
coefficient even for large probabilities $\lambda$ of getting
infected. This low values of $C_{viral}$ justify the assumption made
in our model that the social network is largely irrelevant to
understand the dynamics of information propagation below or even
close to the tipping point.

\section{Viral campaigns general description}
The following describes in some detail the technical and marketing
aspects involved in the execution of the Viral Marketing campaigns
utilized as source of the viral propagation data used in our
studies. It covers 16 different campaigns executed in 11 European
countries, all of them with the same structure, strategy, user
interfaces, data flow or participants conditions.

\medskip

The primary marketing objective of the viral campaign was to
increase the number of subscriptions to the company on-line
newsletter, and the offering consisted in the free subscription to
such newsletter which can be customized according to the
subscriber's interest who was asked to choose from a list of
available generic topics represented by interest codes. The
subscription was formalized by filling in a form located in the main
campaign web page (a.k.a. registration page) of the campaign. A
series of drive-to-web tactics, variable by country, was put in
place to attract visitors to the registration page. This included
e-mail campaigns, banner advertising, search engines placement,
promotion at the company web site and other web based promotional
activities.

\medskip

Additionally, a viral propagation tool consisting of a button
located at the registration page was established to trigger the
message propagation. The caption in that button invited visitors to
recommend the page to friends and colleagues and offered, as
additional incentive for people to forward the page, tickets for a
prize draw to win a laptop computer. Two situations caused
participants to become eligible to receive prize draw tickets:
\begin{itemize}
    \item One ticket was assigned to participants sending any number of
    recommendations to friends or colleagues
    \item Unlimited number of additional tickets were given to the sender for
    each of the recommended friends who would, as a result of such recommendation, subscribe to the newsletter
\end{itemize}

\medskip

The ticket eligibility rules above were designed to discourage
spam-like behavior where recommendations are sent indiscriminately
to individuals not interested in the offering all the while they
encouraged to send the highest possible number of recommendations to
individuals presumed to be interested in the newsletter.
Additionally, the participation rules guarantees that the incentive
was direct consequence of the viral message propagation and not of
registration to the newsletter.

\end{document}